\documentclass[a4paper]{spie}
\usepackage{graphicx}
\usepackage{amsfonts,amssymb,amsbsy,amsmath}

\usepackage{colortbl}

\title{Momentum and rest mass of the covariant state of light\\ in a medium} 
\author{Mikko Partanen and Jukka Tulkki\skiplinehalf
Engineered Nanosystems group, School of Science\\Aalto University, P.O. Box 12200, 00076 Aalto, Finland\\
}

\begin{document} 

\maketitle 

\begin{abstract}
Conventionally, theories of electromagnetic waves in a medium assume that
only the energy of the field propagates in a transparent medium and the medium
is left undisturbed.
Consequently, the transport of mass density and the related
kinetic and elastic energies of atoms is neglected.
We have recently presented foundations
of a covariant theory of light propagation in a medium by considering a light wave simultaneously
with the dynamics of the medium atoms driven by optoelastic forces between the induced dipoles
and the electromagnetic field. In the previously discussed mass-polariton (MP) quasiparticle approach,
we considered the light pulse as an isolated coupled state between the photon
and matter and showed that the momentum and the transferred mass of MP follow unambiguously
from the Lorentz invariance and the fundamental conservation laws of nature. In the present work,
we combine the electrodynamics of continuous media and elasticity theory to account for the space-
and time dependent dynamics of the light pulse and the associated mass and momentum distributions
of the mass density wave (MDW). In this optoelastic continuum dynamics (OCD) approach, we obtain
a numerically accurate solution of the Newtonian continuum
dynamics of the medium when the light pulse is propagating in it. For an incoming Gaussian light pulse
having total energy $E_0$ in vacuum, the OCD simulations of the light pulse propagating in a crystal having
refractive index $n$ give the same momentum $p=nE_0/c$ and the transferred mass $\delta m=(n^2-1)E_0/c^2$
as the MP quasiparticle approach.
Since the elastic forces are included in our theory on
equal footing with the optical forces, our theory also predicts how the mass and thermal
equilibria are re-established by elastic waves.
\end{abstract}
\keywords{mass-polariton, photon momentum, optical forces, electrodynamics, optomechanics}

\section{Introduction}

It is well known in the electrodynamics of continuous media that, when a light
pulse propagates in a medium, the atoms are a subject of field-dipole forces \cite{Landau1984}.
However, the coupled dynamics of the field and the medium driven by the field-dipole
forces has been a subject of very few detailed studies. In this work, we elaborate how
these driving forces give rise to a mass density wave (MDW) in the medium when
a light pulse is propagating in it. We have previously studied MDW in the mass-polariton
(MP) quasiparticle picture \cite{Partanen2017}. In this picture, the coupled state of the field and matter
is considered isolated from the rest of the medium and thus a subject of the covariance
principle and the general conservation laws of nature. The MP quasiparticle picture
also gives a transparent resolution to the centenary Abraham-Minkowski controversy
of the momentum of light in a medium \cite{Cho2010,Leonhardt2006,Pfeifer2007,Barnett2010b,Barnett2010a,Leonhardt2014},
which has also gained much experimental interest
\cite{Campbell2005,Sapiro2009,Jones1954,Jones1978,Walker1975,She2008,Zhang2015}.

\begin{figure}
\centering
 \includegraphics[width=0.68\textwidth]{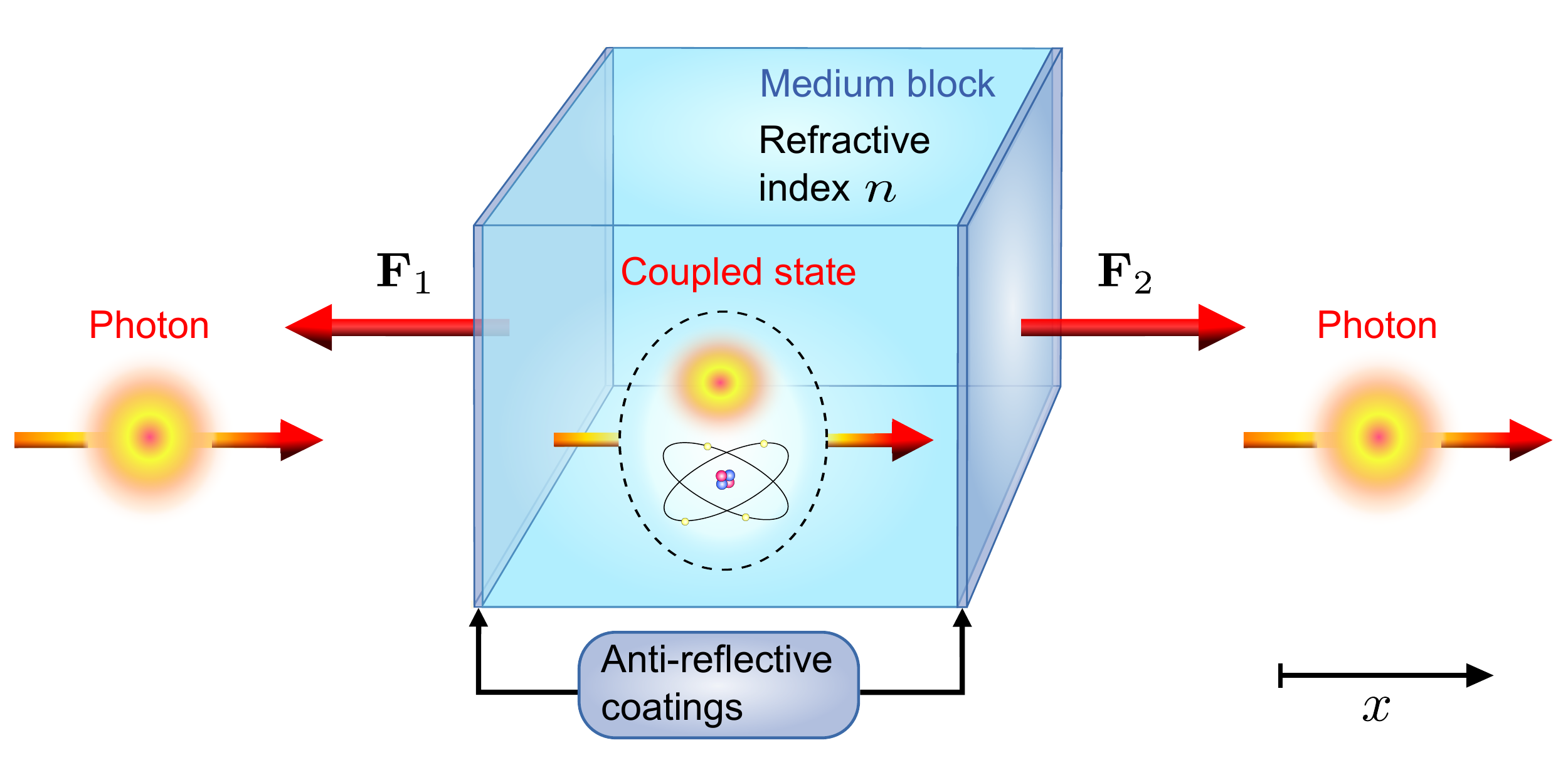}
\caption{\label{fig:illustration}
(Color online) Schematic illustration of a photon propagating
through a transparent medium block with a refractive index $n$. On the left, the photon is incident
from vacuum. Inside the medium block, the photon couples to atoms
forming a quasiparticle, which continues to propagate inside the block.
On the right, the photon continues to propagate in vacuum after the transmission. At the photon entrance
and exit, the medium block experiences recoil forces $\mathbf{F}_1$ and $\mathbf{F}_2$
that depend on the total momentum of light in a medium.
The anti-reflective coatings at the interfaces are included only to simplify the conceptional
understanding of the problem.}
\end{figure}

The problem of light propagation in a medium is depicted in Fig.~\ref{fig:illustration},
which presents the state of the photon before (left) inside the medium (middle)
and after leaving the medium (right). In this work we apply the electrodynamics of
continuous media and continuum mechanics to compute the dynamics of the medium when
a light pulse is propagating in it. In this optoelastic continuum dynamics (OCD)
approach, the total force on a medium element consists of the field-dipole force and
the elastic force resulting from the density variations in the medium. In brief, we
will show that accounting for MDW resulting from the coupled dynamics of the field
and matter allows formulating a fully consistent covariant theory of light propagation
in a medium. OCD theory enables numerical calculation of the mass and momentum distribution
of the atoms moving with the light pulse. We will show that OCD theory provides
an independent but complementary view of how the covariance principle governs the
field-matter coupling in the case of a light pulse propagating in a medium. 

Recently Leonhardt \cite{Leonhardt2014} has applied fluid dynamics to study the momentum
transport of light in fluids. In his work, Lendhardt calculates the force of the field-dipole
interaction on fluid elements. However, Leonhardt does not study the dynamics of the
fluid in the time scale of the oscillating harmonic components of light
as he concentrates on the deformation of the
fluid surface due to a stationary light beam. In the present work, we show that to calculate the
dynamics of the medium one has to account for the exact time dependence of the field induced
force to obtain correct total momentum and transferred mass for the light wave. In addition,
Leonhardt assumes that the fluid is incompressible. This makes the light induced force field to propagate
at infinite speed in the fluid ruining the relativistic invariance and leading to
a fundamentally unsound theory regarding the momentum of light in a medium.
The OCD theory presented in this work can be straightforwardly
generalized to fluids but, in the present simulations, we consider only elastic non-dispersive solids.

\section{Optoelastic continuum dynamics}
\label{sec:ocd}

\subsection{Newton's equation of motion and optoelastic forces}

We will use Newtonian formulation of the continuum mechanics to show
that the optical force gives rise to MDW. Together with the 
associated recoil effect, the MDW effect perturbs the mass density of the medium
from its equilibrium value $\rho_0$. 
In this OCD model, the perturbed atomic mass density of the medium becomes
$\rho_\mathrm{a}(\mathbf{r},t)=\rho_0+\rho_\mathrm{rec}(\mathbf{r},t)+\rho_\text{\tiny MDW}(\mathbf{r},t)$,
where $\rho_\mathrm{rec}(\mathbf{r},t)$
is the mass density perturbance due to the recoil effect and $\rho_\text{\tiny MDW}(\mathbf{r},t)$
is the mass density of MDW.
The mass density perturbances related to the recoil and MDW
effects become spatially separated after the light pulse has penetrated in the medium.
The recoil effect will be shown to lead to the mass density perturbance at the interface while
the MDW follows the light pulse inside the medium.
As the atomic velocities are nonrelativistic in the reference frame where the medium is initially at rest,
the Newtonian mechanics can be applied for the description
of the movement of atoms. Newton's second law for the mass density
of the medium is written as
\begin{equation}
 \rho_\mathrm{a}(\mathbf{r},t)\frac{d^2\boldsymbol{r}_\mathrm{a}(\mathbf{r},t)}{dt^2}=\mathbf{f}_\mathrm{opt}(\mathbf{r},t)+\mathbf{f}_\mathrm{el}(\mathbf{r},t),
 \label{eq:mediumnewton}
\end{equation}
where $\boldsymbol{r}_\mathrm{a}(\mathbf{r},t)$ is the atomic displacement
field of the medium, $\mathbf{f}_\mathrm{opt}(\mathbf{r},t)$ is the optical force density, and
$\mathbf{f}_\mathrm{el}(\mathbf{r},t)$ is the elastic force density.

The optical force density originates from the interaction between the induced dipoles and the
electromagnetic field and it effectively also accounts for the interaction between the induced dipoles.
Using Maxwell's equations and the Lorentz force law,
it can be shown that the optical force density experienced by the induced dipoles
in the medium is given in terms of the electric field $\mathbf{E}(\mathbf{r},t)$ and the Poynting
vector $\mathbf{S}(\mathbf{r},t)$ as \cite{Milonni2010}
\begin{equation}
 \mathbf{f}_\mathrm{opt}(\mathbf{r},t) =-\frac{\varepsilon_0}{2}E(\mathbf{r},t)^2\nabla n^2+\frac{n^2-1}{c^2}\frac{\partial}{\partial t}\mathbf{S}(\mathbf{r},t).
 \label{eq:opticalforcedensity}
\end{equation}

As the atoms are displaced from their equilibrium positions due to the optical force,
they are also affected by the elastic force density $\mathbf{f}_\mathrm{el}(\mathbf{r},t)$
following from Hooke's law.
In the case of a homogeneous isotropic elastic medium,
the stiffness tensor in Hooke's law has only two independent entries.
Typically, these entries are described by using the Lam\'e parameters
or any two independent elastic moduli, such as the bulk modulus $B$
and the shear modulus $G$ \cite{Mavko2003}.
The elastic force density of a homogeneous isotropic medium is then given
in terms of the material displacement field $\mathbf{r}_\mathrm{a}(\mathbf{r},t)$
as \cite{Bedford1994}
\begin{equation}
 \mathbf{f}_\mathrm{el}(\mathbf{r},t)=\textstyle(B+\frac{4}{3}G)\nabla[\nabla\cdot\mathbf{r}_\mathrm{a}(\mathbf{r},t)]-G\nabla\times[\nabla\times\mathbf{r}_\mathrm{a}(\mathbf{r},t)].
 \label{eq:elasticforcedensity}
\end{equation}
The factor $B+\frac{4}{3}G$ in the first term of Eq.~\eqref{eq:elasticforcedensity} is also
known as the P-wave modulus \cite{Mavko2003}.
In the special case of non-viscous fluids, we could set the shear modulus $G$
to zero, in which case the second term of Eq.~\eqref{eq:elasticforcedensity}
becomes zero and the bulk modulus $B$ is the only elastic modulus that remains.

\subsection{Transferred mass and momentum of the mass density wave}

The optical and elastic force densities in Eqs.~\eqref{eq:opticalforcedensity} and \eqref{eq:elasticforcedensity}
and Newton's equation of motion in Eq.~\eqref{eq:mediumnewton}
can be used to simulate the motion of atoms in the medium
as a function of space and time. The total displacement of atoms at position $\mathbf{r}$
is solved from Eq.~\eqref{eq:mediumnewton} by integration as
\begin{equation}
 \mathbf{r}_\mathrm{a}(\mathbf{r},t)=\int_{-\infty}^t\int_{-\infty}^{t''}\frac{d^2\boldsymbol{r}_\mathrm{a}(\mathbf{r},t')}{dt'^2}dt'dt''
 =\int_{-\infty}^t\int_{-\infty}^{t''}\frac{\mathbf{f}_\mathrm{opt}(\mathbf{r},t')+\mathbf{f}_\mathrm{el}(\mathbf{r},t')}{\rho_\mathrm{a}(\mathbf{r},t')}dt'dt''. 
 \label{eq:displacement}
\end{equation}
As the total mass of atoms inside the light pulse is very large when compared to mass equivalent of the field energy,
the perturbed mass density of the medium $\rho_\mathrm{a}(\mathbf{r},t)$
is extremely close to the equilibrium mass density $\rho_0$. Therefore,
when using Eq.~\eqref{eq:displacement}, it is well justified to
approximate the mass density in the denominator of the integrand
with the equilibrium mass density $\rho_0$.

When the light pulse has passed through the medium, the displacement of atoms
at the position $\mathbf{r}$ is given by $\mathbf{r}_\mathrm{a}(\mathbf{r},\infty)$.
We then obtain the displaced volume as
$\delta V=\int\mathbf{r}_\mathrm{a}(\mathbf{r},\infty)\cdot d\mathbf{A}$,
where the integration is performed over the transverse plane
with a surface element vector $d\mathbf{A}$. Using the solution of the atomic displacements
in Eq.~\eqref{eq:displacement}, one obtains an equation for the total transferred mass
$\delta m=\rho_0\delta V$, given by
\begin{equation}
 \delta m=\int\int_{-\infty}^\infty\int_{-\infty}^{t}[\mathbf{f}_\mathrm{opt}(\mathbf{r},t')+\mathbf{f}_\mathrm{el}(\mathbf{r},t')]dt'dt\cdot d\mathbf{A}.
 \label{eq:mdwmass2}
\end{equation}
The total transferred mass $\delta m$ is given in terms of the mass density of MDW as
$\delta m=\int\rho_\text{\tiny MDW}(\mathbf{r},t)dV$. Using Eq.~\eqref{eq:mdwmass2} and the relation $cdt=ndx$,
we then obtain the mass density of MDW, given by
\begin{equation}
 \rho_\text{\tiny MDW}(\mathbf{r},t)=\frac{n}{c}\int_{-\infty}^{t}[\mathbf{f}_\mathrm{opt}(\mathbf{r},t')+\mathbf{f}_\mathrm{el}(\mathbf{r},t')]\cdot\hat{\mathbf{x}}\,dt',
 \label{eq:mdwdensity}
\end{equation}
where $\hat{\mathbf{x}}$ is the unit vector in the direction of light propagation.
With the expressions for the optical and elastic forces in
Eqs.~\eqref{eq:opticalforcedensity} and \eqref{eq:elasticforcedensity}, one can use Eq.~\eqref{eq:mdwdensity}
for numerical simulations of the propagation of light and the associated MDW in the medium.

The velocity distribution of the medium is given by
$\mathbf{v}_\mathrm{a}(\mathbf{r},t)=d\boldsymbol{r}_\mathrm{a}(\mathbf{r},t)/dt$.
Therefore, the momentum of MDW is directly given by
integration of the classical momentum density $\rho_0\mathbf{v}_\mathrm{a}(\mathbf{r},t)$ as
\begin{equation}
 \mathbf{p}_\text{\tiny MDW}=\int \rho_0\mathbf{v}_\mathrm{a}(\mathbf{r},t)d^3r=\int \rho_\text{\tiny MDW}(\mathbf{r},t)\mathbf{v}d^3r,
 \label{eq:mdwmomentum}
\end{equation}
where $\mathbf{v}$ is the velocity vector of MP with length $v=c/n$.
In numerical simulations described in Sec.~\ref{sec:simulations},
it is verified that both forms in Eq.~\eqref{eq:mdwmomentum} give an equal result.

\subsection{Comparison of the OCD and MP quasiparticle approaches}

For the light pulse of energy $E_0$,
the total mass transferred by MDW, given in Eq.\eqref{eq:mdwmass2}, can be also written as
\begin{equation}
 \delta m=\int \rho_\text{\tiny MDW}(\mathbf{r},t)d^3r=(n^2-1)E_0/c^2.
\end{equation}
The right hand side equals the result obtained from the MP quasiparticle model \cite{Partanen2017}.
The total momentum of the coupled state of the field and matter
is a sum of the momenta of the field and MDW, given by
\begin{equation}
 \mathbf{p}_\text{\tiny MP}=\int \rho_0\mathbf{v}_\mathrm{a}(\mathbf{r},t)d^3r+\int \frac{\mathbf{S}(\mathbf{r},t)}{c^2}d^3r=\frac{nE_0}{c}\hat{\mathbf{x}}.
\end{equation}
The first term on the left is the MDW momentum in Eq.~\eqref{eq:mdwmomentum} and
the second term is the momentum density of the electromagnetic field.
On the right, we have the momentum density of the coupled state obtained
from the MP quasiparticle model \cite{Partanen2017}. In the simulations
described in Sec.~\ref{sec:simulations}, it is found that
the OCD and MP quasiparticle model results agree within the numerical
accuracy of the simulations.

\begin{figure}[b]
\centering
\includegraphics[width=0.6\textwidth]{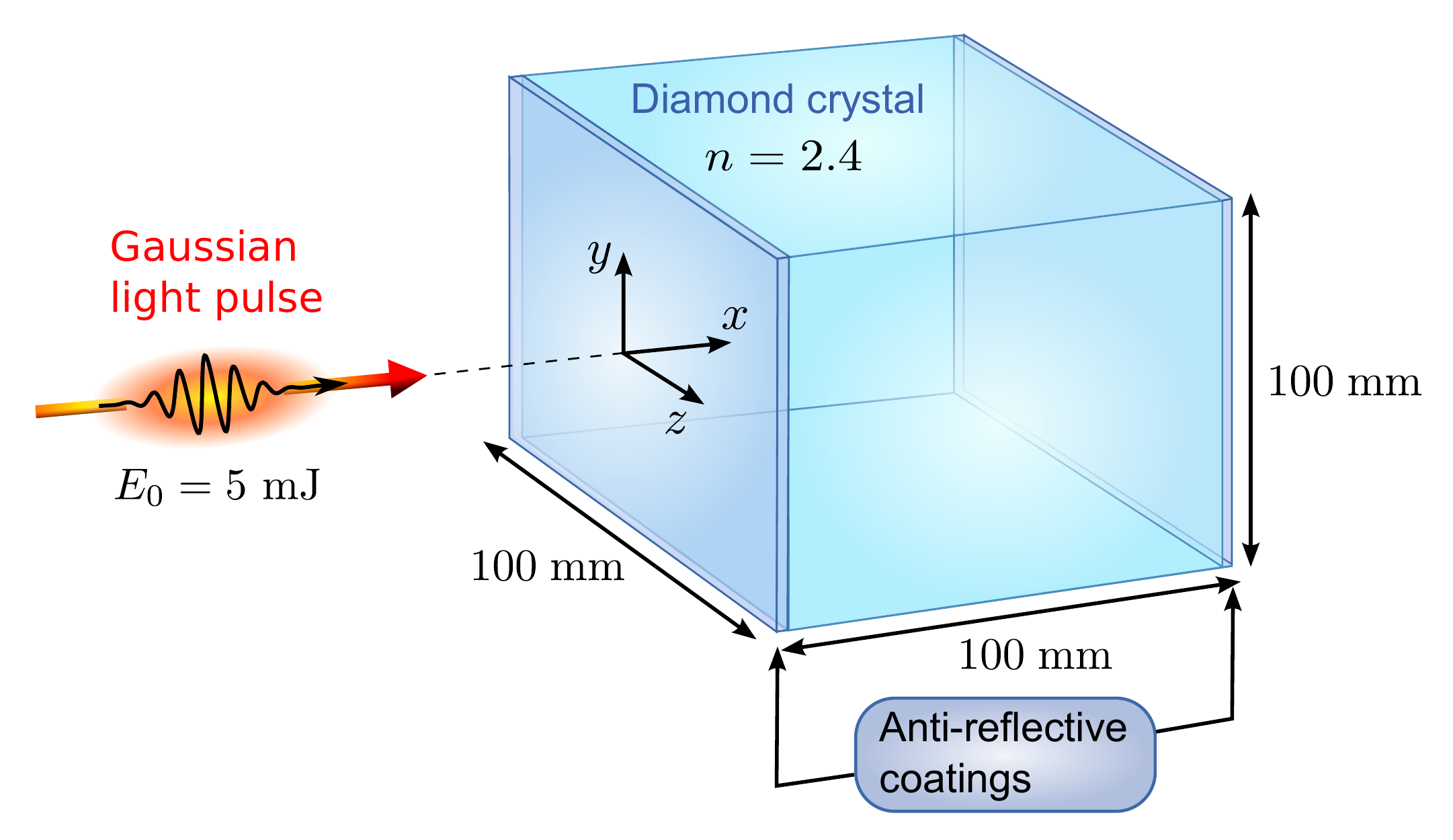}
\caption{\label{fig:simulationgeometry}
(Color online) Illustration of the simulation geometry consisting of a cubic diamond crystal block coated
with anti-reflective coatings. The refractive index of the crystal is $n=2.4$.
A Gaussian light pulse of energy $E_0=5$ mJ propagates in the
direction of the positive $x$-axis and enters the crystal from the left.
The geometry is centered so that the center of the light pulse enters the crystal at $x=y=z=0$.
The first interface of the crystal is located at $x=0$ and the second
interface at $x=100$ mm.}
\end{figure}

\section{Simulations of the mass transfer}
\label{sec:simulations}

We simulate the propagation of a light pulse in the geometry
of a cubic diamond crystal block with anti-reflective coatings
illustrated in Fig.~\ref{fig:simulationgeometry}.
In the $x$ direction, the first and second interfaces of the crystal are located at positions
$x=0$ and $x=100$ mm. In the $y$ and $z$ directions, the geometry is centered so
that the trajectory of the light pulse center follows the line $y=z=0$.
For the diamond crystal, we use a refractive
index $n=2.4$ \cite{Phillip1964}, mass density $\rho_0=3500$ kg/m$^3$ \cite{Lide2004},
bulk modulus $B=443$ GPa \cite{Kittel2005}, and shear modulus $G=478$ GPa \cite{McSkimin1972}.

We assume a titanium-sapphire laser
pulse with a wavelength $\lambda_0=800$ nm in vacuum and
total energy $E_0=5$ mJ.
The wavelength in the diamond crystal is then $\lambda=\lambda_0/n=333$ nm. 
The Gaussian form of the laser pulse is assumed to propagate in the direction
of the positive $x$-axis.
The energy density of the laser pulse averaged over the harmonic cycle is then given by
\begin{equation}
 u(\mathbf{r},t)=E_0\frac{n\Delta k_x\Delta k_y\Delta k_z}{\pi^{3/2}}\,e^{-(n\Delta k_x)^2(x-ct/n)^2}e^{-(\Delta k_y)^2y^2}e^{-(\Delta k_z)^2z^2},
 \label{eq:Gaussianenergydensity}
\end{equation}
where $\Delta k_x$, $\Delta k_y$, and $\Delta k_z$ are the standard deviations
of the wave vector components in vacuum, defining the pulse width
in the $x$, $y$, and $z$ directions.
In the simulations, we use the relative spectral width
of the pulse given by $\Delta\omega/\omega=\Delta k_x/k_0=10^{-5}$. The corresponding standard deviation of position is
$\Delta x=1/(\sqrt{2}\Delta k_x)\approx 9$ mm, and the standard deviation of the
pulse width in time is $\Delta t=n\Delta x/c\approx 30$ ps.
In the transverse direction, we use $\Delta k_y=\Delta k_z=10^{-4} k_0$
corresponding to the standard deviation of position
given by $\Delta y=\Delta z\approx 0.9$ mm.

\subsection{Simulation in one dimension}

In the one-dimensional simulation, the simulation geometry corresponds to a plate that has thickness
$L=100$ mm in the $x$ direction and is infinite in the $y$ and $z$ directions.
The one-dimensional light pulse is Gaussian only in the $x$ direction, which is the direction of propagation.
The one-dimensional pulse is obtained from the three-dimensional pulse in Eq.~\eqref{eq:Gaussianenergydensity}
by dropping out the $y$ and $z$ dependence and renormalizing
the light pulse so that its integral over $x$ gives $E_0/(\lambda/2)^2$.
This corresponds to very high power per unit area, which
was chosen so that we obtain
an order of magnitude estimate of how large atomic displacements
we obtain if the whole vacuum energy $E_0=5$ mJ of the laser pulse can be coupled
to a single mode fiber having a cross section $(\lambda/2)^2$.
The discretization length in the one-dimensional simulation
is $h_x=250$ $\mu$m, which is small compared to the pulse width.

\begin{figure}[b]
\centering
\includegraphics[width=\columnwidth]{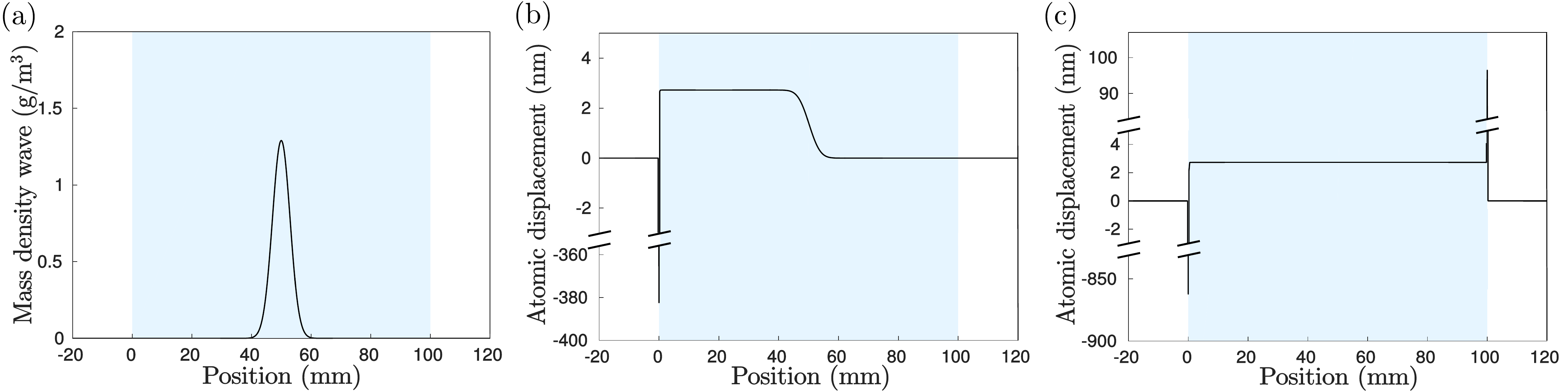}
\caption{\label{fig:mdw}
(Color online) Simulation of the mass transfer in a single-mode diamond waveguide
due to a Gaussian light beam.
(a) The calculated mass density of MDW as a function
of position when the light pulse is in the middle of the crystal.
The light blue background represents the region of the crystal between positions $x=0$ and $x=100$ mm.
(b) The calculated atomic displacements when the light pulse is in the middle of the crystal.
(c) The calculated atomic displacements when the light pulse has just left the crystal.
Note the breaks in the scales of the figures.}
\end{figure}

Figure \ref{fig:mdw}(a) shows the MDW as a function of position
when the light pulse is propagating in the middle of the crystal.
MDW mass density $\rho_\text{\tiny MDW}(\mathbf{r},t)$
is calculated by using Eq.~\eqref{eq:mdwdensity}.
MDW is driven by the optoelastic forces due to the Gaussian
light pulse. The mass density perturbance $\rho_\mathrm{rec}(\mathbf{r},t)$ at the first interface due to the
interface force is not shown in the figure. One can see that
the form of MDW clearly follows the Gaussian
form of the light pulse as expected. When the MDW mass density
in Fig.~\ref{fig:mdw}(a) is integrated over the light pulse, we obtain the total
transferred mass of $2.6\times 10^{-19}$ kg. Dividing this by
the photon number of the light pulse $N_0=E_0/\hbar\omega=2.0\times 10^{16}$, we then obtain the value of
the transferred mass per photon, given by $\delta m=(n^2-1)\hbar\omega/c^2=7.4$ eV/$c^2$,
which corresponds to the value obtained in the MP
quasiparticle approach \cite{Partanen2017}.

Figure \ref{fig:mdw}(b) presents the corresponding atomic displacements
as a function of position.
On the left from the first interface at $x=0$, the atomic
displacement is zero as there are no atoms in vacuum.
As a result of the optoelastic recoil effect described by
the first term of Eq.~\eqref{eq:opticalforcedensity},
a thin material layer at the interface recoils to the left.
In the simulation, the width of the material layer that takes the recoil energy
is the discretization length of $h_x=250$ $\mu$m.
Therefore, we do not account for the atomic scale effects on the refractive
index at the interfaces and the negative atomic displacement at the interface is
only an approximate. Between the positions $x=0$ and $x=40$ mm,
the atomic displacement has a constant value of 
of 2.7 nm. This results from the optical force in the second term
of Eq.~\eqref{eq:opticalforcedensity}.
To the right of the position $x=60$ mm, the atomic displacement is zero
as the leading edge of the light pulse has not yet reached these positions.
In Fig.~\ref{fig:mdw}(b), MDW is manifested by the fact that atoms
are more densely spaced at the position of the light pulse since the atoms
on the left of the pulse have been displaced forward and the atoms
on the right of the pulse are still at their initial positions.

Figure \ref{fig:mdw}(c) shows the atomic displacements just after the light pulse
has left the medium. One can see that all atoms inside the crystal have been displaced
forward from their initial positions. At the both interfaces,
the surface atoms have been displaced outwards from the medium due to the optoelastic recoil effect.
After the transmission of the light pulse, the elastic forces start to restore the mass density equilibrium in the crystal.
Therefore, due to the elastic forces, the magnitudes of the atomic displacements at the interfaces are changing
as a function of time.
When the mass equilibrium has been re-established, the elastic energy that was
left in the crystal, after the transmission of the light pulse, is converted to lattice heat.

\subsection{Simulation in three dimensions}

\begin{figure}
\centering
\includegraphics[width=0.5\textwidth]{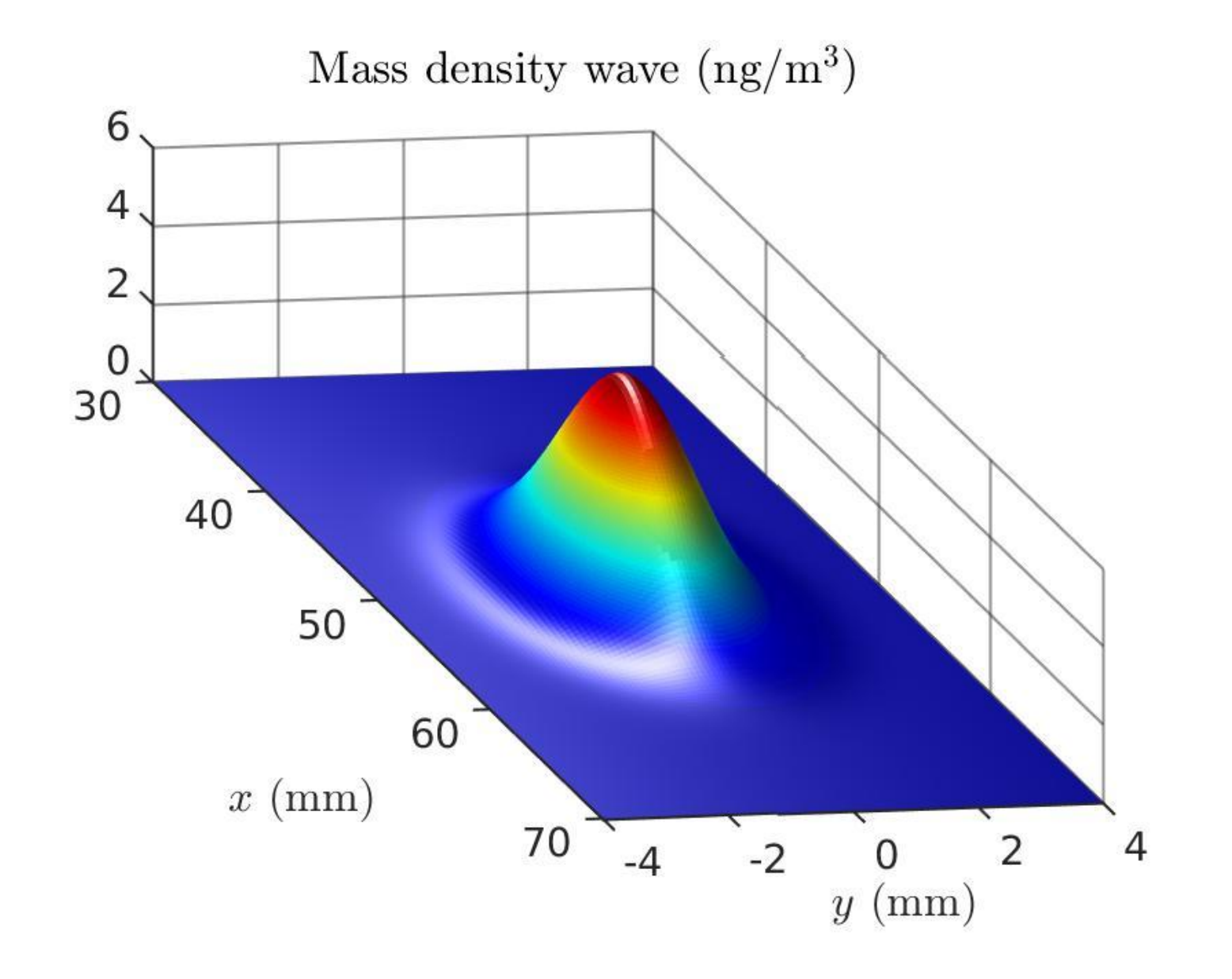}
\caption{\label{fig:mdwsimulation}
(Color online) Simulation of MDW driven by optoelastic forces.
The MDW mass density $\rho_\text{\tiny MDW}(\mathbf{r},t)$
is plotted as a function of position in the plane $z=0$.
The light pulse propagates in the direction of the positive $x$-axis. 
The front end of the crystal is located at $x=0$ and the back end at $x=100$ mm, not
shown in this figure.}
\end{figure}

Figure \ref{fig:mdwsimulation} presents the mass density of MDW simulated in the
full three-dimensional geometry in Fig.~\ref{fig:simulationgeometry}.
The MDW mass density calculated by using Eq.~\eqref{eq:mdwdensity}
is shown as a function of position in the plane $z=0$ when the light pulse is
propagating in the middle of the crystal.
Again, one can see that
the form of MDW clearly follows the Gaussian
form of the light pulse. If the MDW mass density
in Fig.~\ref{fig:mdwsimulation}(a) is integrated over the light pulse, we
obtain the same total transferred mass of $2.6\times 10^{-19}$ kg
as in the case of the one-dimensional simulation.
Therefore, the simulation results are fully consistent with the results
obtained in the MP quasiparticle approach \cite{Partanen2017}.
In the three-dimensional simulations, the maximum atomic displacement
after the light pulse is found to be $1.5\times 10^{-17}$ m,
which is significantly smaller than the value obtained
in the one-dimensional simulation, where the field
is spatially restricted into a smaller volume.

\section{Conclusions}
\label{sec:conclusions}

In conclusion, we have presented a covariant OCD approach for the study
of light propagation in a nondispersive medium,
and compared the OCD approach to the previously discussed MP
quasiparticle picture.
Our analysis shows that, when a photon
enters the crystal, its energy and momentum
are shared by the medium atoms and the
propagating electromagnetic field.
Both the MP quasiparticle picture and the OCD approach
lead to transfer of mass with the light wave.
The transfer of mass with the light wave, in turn, gives rise to nonequilibrium
of the mass density in the medium. When the mass equilibrium is
re-established by relaxation, a small amount of initial photon
energy is converted to lattice heat. These discoveries fundamentally
change our understanding of light-matter interaction.
We have calculated the mass transfer numerically for one- and three-dimensional Gaussian wave packets
and the diamond crystal with realistic material parameters.
The mass transfer and dissipation are
real world phenomena that can also be studied experimentally.
Thus, our work is of great interest to
scientists experimenting with light.

\begin{acknowledgments}
This work has in part been funded by the Academy of Finland and the Aalto Energy Efficiency Research Programme.
\end{acknowledgments}


\end{document}